\documentclass[12pt,a4paper,final]{article}
\usepackage[a4paper,portrait]{geometry}
\usepackage{amsmath,amsfonts,latexsym,amssymb,amsthm}
\usepackage{graphicx,url}

\newcommand{\tr}{{\ensuremath{\mathrm{tr}}}}
\newcommand{\diag}{{\ensuremath{\mathrm{diag}}}}
\newtheorem{thm}{Theorem}
\newtheorem{cor}{Corollary}
\newtheorem{rem}{Remark}

\title{A new method for the estimation of \\
  variance matrix with prescribed zeros \\
  in nonlinear mixed effects models}

\author{Djalil~\textsc{Chafa\"\i} and Didier~\textsc{Concordet}}

\date{August, 2007. Revised March, 2008. \\
 Accepted for publication in Statistics and Computing}

\begin{document}

\maketitle

\begin{abstract}
  We propose a new method for the Maximum Likelihood Estimator (MLE) of
  nonlinear mixed effects models when the variance matrix of Gaussian random
  effects has a prescribed pattern of zeros (PPZ). The method consists of
  coupling the recently developed Iterative Conditional Fitting (ICF)
  algorithm with the Expectation Maximization (EM) algorithm. It provides
  positive definite estimates for any sample size, and does not rely on any
  structural assumption concerning the PPZ. It can be easily adapted to many
  versions of EM.
\end{abstract}


\section{Introduction}

Nonlinear mixed effects models are widely used in population
Pharmacology for Pharmacokinetics \& Pharmacodynamics (PK/PD)
modelling. Such models can be seen as special cases of repeated
measurement data, for which the asymptotics concern the number of
individuals rather than the number of measures per individual, see
for instance \cite{davidian} and references therein. In this
article, we show that for nonlinear mixed effects models with
Gaussian random effects, Expectation Maximization (EM) like
algorithms for the computation of the Maximum Likelihood Estimator
(MLE) can be coupled with Iterative Conditional Fitting (ICF) like
algorithms in order to take into account a prescribed pattern of
zeros (PPZ) in the variance matrix of the random effect. The ICF
algorithm has been developed very recently \cite{chaudhuri} in the
context of directly observed Gaussian graphical models. Finding an
adequate approach for generic PPZ in the context of nonlinear mixed
effects models is a long standing problem. Our approach provides a
true solution for the M step of EM in this context, for any PPZ. It
is thus far more satisfactory than the standard approaches used in
the existing software packages such as NLMIXED (SAS), NONMEM, nlme
(S-Plus and GNU-R), or Monolix.

For instance, a traditional model used in population PK/PD is of the form
\begin{equation}\label{eq:mod1}
    Y_{i}=F(X_{i})+g(X_{i},\theta)\varepsilon_{i},\quad 1\leq i\leq N,
\end{equation}
where $Y_{i}$ is the vector of concentrations/effects observed on the $i^{th}$
individual for the drug of interest. Here $F$ is a known function, often
nonlinear. The $q$-vectors $X_{i}$ represent the unobserved individual
parameters assumed independent and identically distributed $\mathcal{N}\left
  (m,\Sigma\right )$, the $\varepsilon_{i}$ are $\mathcal{N}(0,I)$ ,
unobserved, independent of the $X_{i}$ and independent. The matrix
$g(X_{i},\theta)$ is the Cholesky transform of a positive definite matrix that
depends on a parameter $\theta\in\Theta\subset\mathbb{R}^{p}.$

\begin{figure}[b]
  \begin{center}
    \includegraphics[scale=0.5,angle=-90]{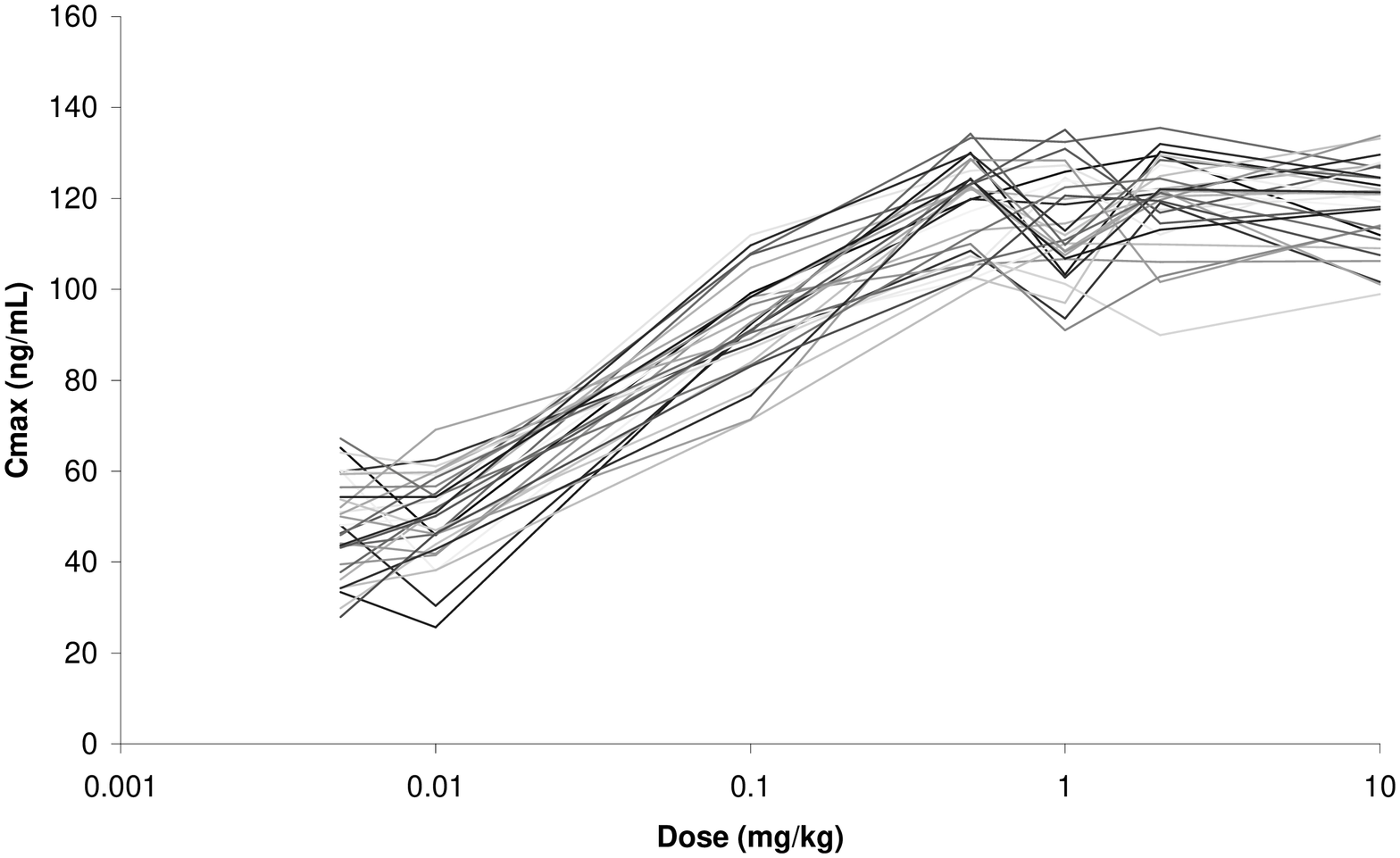}
    \caption{Maximum cortisol concentrations observed after IV
      administrations of ACTH in 30 horses.}
    \label{fig:cortisol}
  \end{center}
\end{figure}

For instance, Figure \ref{fig:cortisol} represents the maximum
concentrations of cortisol ($Y_{i}$) obtained after giving fixed
doses of ACTH to $N=30$ horses. Each individual has its own curve
described by the parameter $X_{i}$. The $g(X_{i},\theta)$ matrix
defines the variance heterogeneity of concentrations obtained with
different doses. This matrix is often assumed to be equal to $\theta
\diag(|F(X_{i}|).$ The reader will later find a study of a model
like \eqref{eq:mod1} for this cortisol data set.

One of the main goals in population PK/PD is to describe the
distribution of the $X_{i}$'s by observation of the $Y_{i}$'s. This
amounts to the estimation of $(m,\Sigma,\theta)$ from the $Y_{i}$'s.
Recall that the $X_i$'s are not observed. A natural approach is to
compute the MLE by maximizing
\begin{equation}\label{eq:like}
  L(m,\Sigma,\theta) %
  = \prod_{i=1}^{N}\int \frac{1}{|g(x_{i},\theta)|}\phi\left (g^{-1}(x_{i},\theta)(Y_{i}-F(x_{i}))\right ) \phi_{m,\Sigma}(x_{i}) dx_{i}
\end{equation}
where $\phi_{m,\Sigma}(.)$ is the probability density function of the
$\mathcal{N}\left (m,\Sigma\right )$ distribution and $|g(x_{i},\theta)|$
denotes the determinant of the matrix $|g(x_{i},\theta)|$. Except for specific
models, such as \emph{Gaussian linear} mixed effects models, maximum
likelihood estimators have no closed form. Several methods have been proposed
for estimation of the parameter $(m,\Sigma,\theta)$ in these models. The
methods suggested by Beal and Sheiner \cite{beal} or Lindstrom and Bates
\cite{lindstrom} are based on a linearization of the conditional model
\eqref{eq:mod1} with respect to the vector $X_{i}$ about $0$ or about a
posterior mode. Pinheiro and Bates \cite{pinheiro}, Vonesh and Carter
\cite{vonesh}, and Wolfinger \cite{wolfinger} proposed Laplacian
approximations of the likelihood. Importance sampling approximations
\cite{pinheiro}, Gaussian quadratures \cite{davidian}, and pseudo-likelihood
methods \cite{concordet} have also been investigated. The reader will find a
detailed analysis of these methods in the book by Davidian and Giltinan
\cite{davidian}. More recently, stochastic versions of the EM algorithm have
been proposed, see for instance \cite{lavielle} and \cite{wang}. These EM like
algorithms converge to the MLE under some regularity and identifiability
conditions.

In many real situations, the kineticist's knowledge of the drug
mechanism imposes a specific independence pattern on some components
of $X_{i}$. This means that the variance matrix $\Sigma$ contains a
PPZ. The estimation of $(m,\Sigma,\theta)$ in the presence of a PPZ
in $\Sigma$ is problematic due to the positive definiteness
constraint in the optimization. Pinheiro and Bates \cite{pinheiro}
studied different parameterizations of $\Sigma$ that ensure the
definite positiveness of the estimate. In particular they suggested
the usage of a Cholesky like parameterization. Unfortunately, except
for the case where $\Sigma$ is a block diagonal matrix up to
coordinates permutation, Cholesky like parametrizations do not
preserve the structure of the PPZ and are thus useless. Kuhn and
Lavielle proposed estimating $\Sigma$ in two steps in the
implementation of their EM like algorithm. First, $\Sigma$ is
estimated without any constraint, then zeros are plugged according
to the PPZ into the estimate provided by the first step. This method
is widely used in practice. Unfortunately, by ``forcing the zeros''
in this way, nothing guarantees that the obtained estimate is still
a positive matrix, and even when it is positive definite, it is not
the maximum likelihood in general.

For \emph{Gaussian linear} mixed effects models, the algorithm of
Anderson \cite{anderson} deals with any linear hypothesis on the
variance matrix of the random effect (a PPZ for instance).
Unfortunately, the estimate is not necessarily positive definite,
see for instance \cite{chaudhuri}. To our knowledge, no method is
available for MLE of \emph{nonlinear} models such as \eqref{eq:mod1}
when the variance matrix $\Sigma$ of the random effect has a PPZ.

The aim of this article is to propose a general method for the
estimation of $(m,\Sigma,\theta)$ in the presence of a PPZ in
$\Sigma$. The method uses the ICF algorithm to perform the
Maximization step of the EM algorithm. In other words, we couple EM
and ICF in order to compute the MLE (or at least a stationary point
of the likelihood) of $(m,\Sigma,\theta)$ when $\Sigma$ has a PPZ.

The ICF algorithm was developed recently by Chaudhuri \emph{et al.}
in \cite{chaudhuri} to estimate a variance matrix with PPZ of
\emph{observed Gaussian} random variables. In contrast, the random
effects $X_i$'s in \eqref{eq:mod1} are Gaussian but \emph{not
observed}, and that is why we couple ICF with EM. The ICF converges
towards positive definite saddle-points or local maxima of the
likelihood function irrespective of the PPZ. The idea behind ICF is
not new in the framework of graphical models, and is inspired by the
famous Iterative Proportional Fitting (IPF) algorithm. We refer to
\cite{chaudhuri} for a review. Some alternative algorithms to ICF
are available for specific PPZ, such as chain graph models
\cite{cox} or non-chordal graph models \cite{dahl}. The ICF
algorithm is attractive because it does not rely on a specific
structure of the PPZ.

The rest of the article is organized as follows. In section 2, we
give some of the properties and drawbacks of the popular ``zero
forced'' estimator, that consists of plugging zeros according to the
PPZ into a full variance matrix. In section 3, we recall the main
properties of the EM algorithm for models such as \eqref{eq:mod1}.
Section 4 is devoted to the ICF algorithm, and to the coupling of
ICF with EM. Section 5 contains the step by step analysis of a model
like \eqref{eq:mod1} for the cortisol data set depicted in Figure
\ref{fig:cortisol}. In the last section, we perform a simulation
study that quantifies the benefit of our EM+ICF approach on the
model used for the cortisol data set.

\section{The Zero forced estimator}

Assume for example that for some specific model we get the following MLE
for~$\Sigma$:
$$
\widehat\Sigma_{uc}=\left(
  \begin{array}{rrr}
    4&-3 & 3 \\
    -3 & 4 & -3 \\
    3 & -3 & 4 \\
  \end{array}
\right),
$$
without taking into account the PPZ in $\Sigma$. We will refer to
this as the \emph{unconstrained} estimation. If the PPZ consists of
$\Sigma_{13}=\Sigma_{31}=0$, the ``zero forced'' estimation of
$\Sigma$ is simply given by
$$
\widehat\Sigma_{zf}=\left(
  \begin{array}{rrr}
    4&-3 & 0 \\
    -3 & 4 & -3 \\
    0 & -3 & 4 \\
  \end{array}
\right).
$$
The unconstrained estimate $\widehat\Sigma_{uc}$ is a positive
definite matrix but the ``zero forced'' estimate
$\widehat\Sigma_{zf}$ is not. However we know that for a regular
model, the unconstrained MLE is consistent. Therefore
$\widehat\Sigma_{uc}$ converges componentwise towards the true
matrix $\Sigma$ with PPZ. Consequently, there exists a random sample
size from which the ``zero forced'' estimator is a positive definite
matrix but this sample size is somewhat difficult to obtain.

A possible ploy allowing to build a positive definite consistent
estimator of $\Sigma$ could be as follows. Compute the unconstrained
estimator and denote it by $\widehat\Sigma_{zf}$, the corresponding
``zero forced'' estimator. Nothing guarantees that its lower
eigenvalue $\lambda_{min}$ is positive but since
$\widehat\Sigma_{zf}$ is a consistent estimator of $\Sigma$, the
quantity
$$
(\lambda_{min})_{-}\triangleq\max\{-\lambda_{min},0\}
$$
is a random sequence of positive numbers that converges almost-surely to zero.
Now, consider some auxiliary sequence of positive real numbers $(u_{N})$ that
goes to zero with the sample size $N$ (e.g. $U_{N}=1/N^{2}$), then, for any
sample size $N$, the matrix
$$
\widehat\Sigma_{zf}+\left ((\lambda_{min})_{-}+u_{N}\right )I
$$
is a positive definite consistent estimator of $\Sigma$, and
features the same PPZ. Its main drawback is that its diagonal terms
are biased and that the choice of the $(u_{N})$ sequence is
arbitrary. A better way to proceed is to directly consider the MLE
of $\Sigma$ with PPZ, which is precisely our aim in the next
sections.

\section{The EM algorithm}

The EM algorithm \cite{dempster} is a popular method to estimate
parameters of a model with non-observed or incomplete data. Let us
briefly recall how its general form works as introduced by Dempster
et al. The EM algorithm consists of iterations of an Expectation and
a Maximization step. At the $k^{th}$ iteration, the E step computes
the conditional expectation of the log-likelihood of the complete
data $(Y,X)$ with respect to the distribution of the missing, or
non-observed, data $X$ given the observed data $Y$ at the current
estimated parameter value $\psi^{(k)}$:
$$ Q\left (\psi, \psi^{(k)}\right )=E\left [\log P(Y,X)|Y, \psi^{(k)}\right ].$$
The M step finds $\psi^{(k+1)}$ so that for all $\psi$ in the
parameter space $\Psi$
$$ \psi^{(k+1)}=\arg\sup_{\psi\in\Psi} Q\left (\psi, \psi^{(k)}\right
).$$ These two-step iterations are repeated until convergence. The
essential property of the EM algorithm is that the likelihood
increases monotonically along the iterations. Under some
identifiability and regularity conditions, this algorithm converges
to a stationary point of the likelihood, see for instance \cite{wu}.

Let us now describe more precisely this algorithm for model \eqref{eq:mod1}.
We need first to define the parameter space on which the M step is to be
performed. In this model, the parameter to be estimated is $\psi=\left (m,
  \Sigma, \theta\right )$. The variance matrix $\Sigma$ lives in a subset of
the set $S_{q}^{+}$ of $q\times q$ symmetric positive definite matrices. More
precisely, let $\Pi$ be the set of subsets of $\left \{ (i,j); 1\leq i<j\leq
  q\right \}$. For any $\pi\in\Pi$, the set
$$
S_{q}^{+}(\pi)\triangleq\{A\in S_{q}^{+}; \forall (i,j)\in\pi, A_{ij}=0\}
$$
is formed by the symmetric positive definite matrices that have zeros located
in $\pi$. The PPZ in $\Sigma$ is represented by an element $\pi$ of $\Pi$. We
thus assume that for some $\pi\in\Pi$, $\psi=\left (m, \Sigma, \theta\right
)\in \Psi\triangleq M\times S_{q}^{+}(\pi)\times \Theta$ where $M$ and
$\Theta$ are open subsets of $\mathbb{R}^{q}$ and $\mathbb{R}^{p}$
respectively.

At the $k^{th}$ iteration the Expectation step consists of the
computation of
$$Q\left (m,\Sigma,\theta|m_{k},\Sigma_{k},\theta_{k}\right )=\sum_{i}
\mathbf{E}\left (\log \frac{1}{|g(X_{i},\theta)|}\phi\left
(g^{-1}(X_{i},\theta)(Y_{i}-F(X_{i}))\right ) \phi_{m,\Sigma}(X_{i})
|Y_{i},\Sigma_{k},m_{k},\theta_{k}\right )$$ where
$$\mathbf{E}\left ( f(X,Y)|Y,\Sigma,m,\theta\right )\triangleq \int f(x,Y)
\frac{\phi\left (g^{-1}(x,\theta)(Y-F(x))\right ) \phi_{m,\Sigma}(x)
}{|g(x,\theta)|\int \phi\left (g^{-1}(u,\theta)(Y-F(u))\right
)/|g(u,\theta)| \phi_{m,\Sigma}(u) du}dx.$$

The maximization step computes
$$\left (m_{k+1},\Sigma_{k+1},\theta_{k+1}\right )=\arg\sup_{M\times S_{q}^{+}(\pi)\times \Theta}Q\left (m,\Sigma,\theta|m_{k},\Sigma_{k},\theta_{k}\right ).$$
For model \eqref{eq:mod1}, the integral that appears in the E step
can be split into two parts . The E step reduces to calculate
$$Q\left (m,\Sigma,\theta|m_{k},\Sigma_{k},\theta_{k}\right )=Q_{1}\left (m,\Sigma|m_{k},\Sigma_{k},\theta_{k}\right )
+Q_{2}\left (\theta|m_{k},\Sigma_{k},\theta_{k}\right )$$ where
\begin{align*}
  Q_{1}\left (m,\Sigma|Y_{i},m_{k},\Sigma_{k},\theta_{k}\right )
  &\triangleq  \sum_{i}\mathbf{E}\left (\log \phi_{m,\Sigma}(X_{i})|Y_{i},m_{k},\Sigma_{k}\right )\\
  &=-\frac{1}{2}\sum_{i}\mathbf{E}\left (
    \left (X_{i}-m\right )'\Sigma^{-1} \left (X_{i}-m\right )|Y_{i},m_{k},\Sigma_{k}\right )-\frac{N}{2}\log |\Sigma|\\
  &=-\frac{N}{2}\tr\left(\frac{1}{N}\sum_{i}\mathbf{E}\left(\left(X_{i}-m\right)\left (X_{i}-m\right)'|Y_{i},m_{k},\Sigma_{k}\right) \Sigma^{-1}\right) \\
    &\quad -\frac{N}{2}\log|\Sigma|,
\end{align*}
and
$$
Q_{2}\left (\theta|m_{k},\Sigma_{k},\theta_{k}\right ) \triangleq
\sum_{i} \mathbf{E}\left (\log \phi\left
(g^{-1}(X_{i},\theta)(Y_{i}-F(X_{i}))\right
)-\log(|g(X_{i},\theta)|)|Y_{i},m_{k},\Sigma_{k},\theta_{k}\right
).$$ It follows that the M step can also be decomposed into two
parts
$$
\sup_{M\times S_{q}^{+}(\pi)\times \Theta}Q\left
  (m,\Sigma,\theta|m_{k},\Sigma_{k},\theta_{k}\right ) = \sup_{M\times
  S_{q}^{+}(\pi)}Q_{1}\left (m,\Sigma|m_{k},\Sigma_{k},\theta_{k}\right )+
\sup_{\Theta}Q_{2}\left (\theta|m_{k},\Sigma_{k},\theta_{k}\right ).
$$
Note that in the $M$ step the maximization with respect to
$(m,\Sigma)$ is separated from that of $\theta$. The function
$Q_{2}$ depends only on $\theta$ via $g$.

\begin{rem}
  In most applications, $h$ is the probability density function of a
  standard Gaussian distribution, and $\theta$ is a variance matrix that can possibly
  contain a PPZ. In that case, its maximization can be performed using the ICF
  method, as for $\Sigma$, as described hereafter.
\end{rem}

Maximization of $Q_{1}$ leads to
\begin{equation}\label{eq:moyenne}
m_{k+1}=\frac{1}{N}\sum_{i}  \mathbf{E}\left ( X_{i}|Y_{i},\Sigma_{k},m_{k},\theta_{k}\right )
\end{equation} and
\begin{equation}\label{eq:opti1}
\Sigma_{k+1}=\arg\inf_{\Sigma\in S_{q}^{+}(\pi)}\tr \left ( {\tilde
X} \Sigma^{-1}\right )+\log |\Sigma|
\end{equation}
where
\begin{equation}\label{eq:variance}
{\tilde X}=\frac{1}{N}\sum_{i}\mathbf{E}\left (\left ( X_{i}-m_{k+1}\right )\left ( X_{i}-m_{k+1}\right )'|Y_{i},\Sigma_{k},m_{k},\theta_{k}\right ).
\end{equation}
The matrix ${\tilde X}$ is an empirical conditional variance matrix.
The random vectors $X_{i}$ are not observed, however, the matrix
${\tilde X}$ can be evaluated at each iteration of the algorithm.
When $\pi=\emptyset$, that is when $\Sigma$ has no PPZ,
$\Sigma_{k+1}$ reduces to ${\tilde X}$. When $\pi\neq\emptyset$ the
maximum of $Q_{1}(\Sigma)$ must be sought in $S_{q}^{+}(\pi)$. The
next section deals with this problem.

\section{Estimating the variance matrix with ICF}

We have seen in the previous section that the M step of the EM
algorithm involves a maximization problem such as
\begin{equation}\label{eq:opti}
  \Sigma_{k+1}=\arg\inf_{\Sigma\in S_{q}^{+}(\pi)}\tr \left ( {\tilde X} \Sigma^{-1}\right )+\log |\Sigma|.
\end{equation}
The difficulty here is that the optimization is not performed on the
entire cone of symmetric definite matrices but only on a sub-cone
that contains the matrices with PPZ. It is clear that a standard
gradient-like algorithm does not fit these constraints. The usual
method to get rid of the definite-positiveness constraint is to use
a Cholesky like decomposition. Unfortunately, these decompositions
do not preserve the PPZ when $\Sigma$ is not a block diagonal matrix
up to a permutation of the coordinates. The algorithm described
hereafter allows one to move within $S_{q}^{+}(\pi)$ whatever $\pi$
may be.

First, note that in the absence of PPZ in $\Sigma$, \emph{i.e.} when
$\pi=\emptyset$, the solution of (\ref{eq:opti}) is
$\Sigma_{k+1}={\tilde X}$. We assume from now on that
$\pi\neq\emptyset.$ The case $q = 2$ is trivial since the only
possible zero is $\Sigma_{12}=0$. In this case, the ``zero forced''
estimator is always positive definite and is the solution of
(\ref{eq:opti}). We assume in the sequel that $q >2$ and that
$\pi\neq \emptyset$.

The method we propose is the core of the ICF algorithm presented in
\cite{chaudhuri}. Even if Chaudhuri \& al. did not express it at
such, it is mainly based on the specific properties of the Schur
complement of a matrix. Let us recall the following classical
result, which can be found in \cite{MR1084815} or \cite{MR2160825}
for instance.
\begin{thm}[Schur]
  Let $l$ be an integer in $\{1,\ldots, q\}$. Consider two vectors $U$ and $V$
  that respectively belong to $\mathbb{R}^{q-l}$ and $\mathbb{R}^{l}$, the
  q-vector $Z'=(U',V')$ and a matrix $\Sigma\in S^{+}_{q}$ that
  admits the following block decomposition
  \begin{equation*}
    \Sigma=\left (%
      \begin{array}{cc}
        A & B\\
        B' & C\\
      \end{array}%
    \right )\\
  \end{equation*}
  where $A\in S^{+}_{q-l}$, $B$ is a $(q-l)\times l$ matrix and $C\in
  S^{+}_{l}$. The Schur complement of $A$ is the matrix $S\triangleq
  C-A'^{-1}B.$ It belongs to $S^{+}_{l}$ and moreover
  \begin{itemize}
  \item[i)] $\det(\Sigma) =\det(A) \det(S)$,
  \item[ii)] $Z'\Sigma^{-1}Z=U' A^{-1}U+
    \left (V-B'A^{-1}U\right )'S^{-1}\left
      (V-B'A^{-1}U\right ).$
  \end{itemize}
\end{thm}

The Schur complement appears naturally when the random vector $Z$
described in the previous Theorem is distributed according to
$\mathcal{N}\left (0,
  \Sigma\right )$. More precisely we have:
$$\mathcal{L}\left (V|U\right
)=\mathcal{N}(B'A^{-1}U,S).$$

Note that properties i) and ii) remain true after permutation of the
rows of $\Sigma$. In the particular case where $l=1$, we can easily
derive the following property.

\begin{cor}
  We keep the same notations as in the previous proposition. For any
  $j\in\{1\ldots q\}$, let $A=\Sigma_{-j,-j}$ be the submatrix of $\Sigma$
  obtained by removing its $j^{th}$ row and column , $B=\Sigma_{-j,j}$ the
  $j^{th}$ column vector of $\Sigma$ in which the $j^{th}$ row has been
  removed, $C=\Sigma_{j,j}$. The column vector $U$ and the positive real
  number $V$ are respectively obtained by removing the $j^{th}$ row of $Z$ and
  as $Z_{j}$. Then, using these notations, the Schur complement of
  $A=\Sigma_{-j,-j}$ is the real positive number $S$ given by the previous
  proposition and properties i) and ii) hold.
\end{cor}
We are now able to solve the optimization problem (\ref{eq:opti1})
by running iteratively the decomposition of the Corollary over the
columns of $\Sigma$. Set $T_{i} =X_{i}-m_{k+1}$ and note that from
(\ref{eq:opti}) the function that has to be minimized can be
rewritten as
\begin{equation}\label{eq:vrais}
K(\Sigma,T_{1},\ldots,T_{N}) = \frac{1}{N} \sum_{i}\mathbf{E}\left (
T_{i}'\Sigma^{-1} T_{i}|Y_{i},m_{k},\Sigma_{k}\right )+\log
|\Sigma|.
\end{equation}

For the $j^{th}$ column of $\Sigma$ we set $A=\Sigma_{-j,-j}$,
$B=\Sigma_{-j,j}$, $C=\Sigma_{j,j}$ and $S=C-B'A^{-1}B$ so that we
can now write (\ref{eq:vrais}) as
\begin{equation*}
K(\Sigma,T_{1},\ldots,T_{N}) = K(A,U_{1},\ldots,U_{N})+
K(S,V_{1}-B'A^{-1}U_{1},\ldots,V_{N}-B'A^{-1}U_{N})
\end{equation*}
where $V_{i}$ is the $j^{th}$ component of $T_{i}$ and $U_{i}$ is
obtained by removing the $j^{th}$ component of $T_{i}$. Therefore,
if $A$ is fixed, the partial optimization of
$K(\Sigma,T_{1},\ldots,T_{N})$ with respect to $(B,S)$ can be
reduced to the global optimization of
\begin{eqnarray*}
  K(S,V_{1}-B'A^{-1}U_{1}&,&\ldots,V_{N}-B'A^{-1}U_{N})=\\
  &{}&\frac{1}{NS}\sum_{i}\mathbf{E}\left (\left
      (V_{i}-B'A^{-1}U_{i}\right )^{2}|Y_{i},m_{k},\Sigma_{k}\right )+\log (S)
\end{eqnarray*}
which is a standard least-squares problem. The optimization with
respect to $B$ and $S$ leads to
\begin{eqnarray}\label{eq:bopt}
  B_{opt}&=&\left [\sum_{i}\mathbf{E}\left (\left ((A^{-1}U_{i})(A^{-1}U_{i})'\right )|Y_{i},m_{k},\Sigma_{k}\right )\right ]^{-1}
  \sum_{i}\mathbf{E}\left (V_{i}A^{-1}U_{i}|Y_{i},m_{k},\Sigma_{k}\right )\\
  &=&A\left [\sum_{i}\mathbf{E}\left (U_{i}U_{i}'|Y_{i},m_{k},\Sigma_{k}\right )\right ]^{-1}
  \sum_{i}\mathbf{E}\left (V_{i}U_{i}|Y_{i},m_{k},\Sigma_{k}\right ).
\end{eqnarray}
and
\begin{eqnarray}\label{eq:sopt}S_{opt}=\frac{1}{N}\sum_{i}\mathbf{E}\left
    (\left
      (V_{i}-B_{opt}'A^{-1}U_{i}\right )^{2}|Y_{i},m_{k},\Sigma_{k}\right
  ).
\end{eqnarray}
The vector $B=\Sigma_{-j,j}$ may contain some PPZ. These components
are not optimized and are thus left at zero. This only decreases the
dimension of the optimization problem. We deduce that after this
step on the $j^{th}$ column of $\Sigma$, $C=\Sigma_{j,j}$ and
$B=\Sigma_{-j,j}$ must be respectively updated with
$$
\Sigma_{j,j}^{new}=S_{opt}+B_{opt}'\left (\Sigma_{-j,-j}\right
)^{-1}B_{opt} \quad\text{and}\quad \Sigma_{-j,j}^{new}=B_{opt}.
$$
The striking property of this step is that it allows us to move
within $S^{+}_{q}$ without affecting the prescribed null components
of $\Sigma$: $A=\Sigma_{-j,-j}$ and the null components of
$B=\Sigma_{-j,j}$ are left unchanged. Since $\Sigma$ is assumed
positive definite $C=\Sigma_{j,j}$ cannot be zero.

As already mentioned, iterations of these steps converge to a local
maximum of $K(\Sigma,T_{1},\ldots,T_{N})$, see for instance
\cite{chaudhuri}.

\section{The cortisol data set}

In this section, we use a practical example to illustrate the
implementation of an EM algorithm coupled with the ICF algorithm. In
order to explore the endocrine function of horse, a sample of horses
($N=30$) was given eight doses of ACTH by intravenous route. The
ACTH stimulates the adrenal gland that produces cortisol. The
concentration profiles of cortisol in plasma were summarized by the
maximal concentration reached after the ACTH administration (see
Figure \ref{fig:cortisol}).
%

The seven doses of ACTH given to each animal were (in mg/kg)
$$
0.005,\quad 0.01,\quad 0.1,\quad 0.5,\quad 1,\quad 2,\quad 10.
$$
The production of cortisol is modelled as
\begin{equation}\label{eq:mod2}
  Y_{ij}=
  \left (X_{1i}+\frac{X_{2i}d_{j}^{X_{3i}}}{X_{4i}^{X_{3i}}+d_{j}^{X_{3i}}}\right )\left (1+\sigma\varepsilon_{ij}\right ),
  \quad 1\leq j\leq 7,\quad 1\leq i\leq 30,
\end{equation}
where $Y_{ij}$ is the maximal cortisol concentration observed in the $i^{th}$
horse after administration of a dose $d_{j}$ of ACTH, $X_{i}'=\left
  (X_{i1},\ldots, X_{i4}\right )$ is a random vector that contains the
individual parameters for the $i^{th}$ animal. We assume that the random
vectors $X_{i}$ are independent and identically distributed $\mathcal{N}\left
  (m,\Sigma\right )$ and that the residual terms $\varepsilon_{i}'=\left
  (\varepsilon_{i1},\ldots, \varepsilon_{i7}\right )$ are independent and
identically distributed $\mathcal{N}\left (0,I_{7}\right )$.
Moreover, the $X_{i}$'s and $\varepsilon_{i}$'s are assumed to be
mutually independent. In this example, $p=4$, $\theta=\sigma^{2}$
and
$$
g(X,\theta) %
=\sigma \diag\left(X_{1}%
  +\frac{X_{2}d_{j}^{X_{3}}}{X_{4i}^{X_{3}}+d_{j}^{X_{3}}}\right)_{j=1\ldots 7}.
$$
According to the kineticist, the correlations between $X_{i1}$ and $X_{i4}$
and $X_{i3}$ and $X_{i4}$ should be zero and thus $\Sigma$ has the following
structure
$$
\Sigma=\left(
  \begin{array}{cccc}
    . & . & . & 0 \\
    . & . & . & . \\
    . & . & . & 0 \\
    0 & . & 0 & . \\
  \end{array}
\right).
$$
In some problems, no \emph{a priori} information is available for
the possible zero correlation between the components of $X_{i}$. The
method of multiple testing of correlation, as described in Drton and
Perlman \cite{drton}, may be used in such cases to reveal the
structure of $\Sigma$.

The estimation of model parameters requires evaluation of the
conditional expectations of functions such as  $\mathbf{E}\left (
  f(X_{i},Y_{i})|Y_{i},\Sigma_{k},m_{k},\sigma^{2}_{k}\right ).$ A standard
approach is to use a stochastic version of EM that consists of the
simulation of a Markov Chain $(X^{(l)}_{i})_{l}$ with $P\left
  (.|Y_{i},\Sigma_{k},m_{k},\sigma^{2}_{k}\right )$ as unique stationary
distribution by using a Metropolis-Hastings algorithm and the
approximation of the conditional expectation by
$$
\mathbf{E} %
\left( f(X_{i},Y_{i})|Y_{i},\Sigma_{k},m_{k},\sigma^{2}_{k}\right) %
\approx\frac{1}{L}\sum_{l=1}^{l}f(X^{(l)}_{i},Y_{i}).
$$
For the analysis of the cortisol data we chose $L=500.$ We simulated the
Markov chain with the Metropolis-Hastings algorithm with $\mathcal{N}\left
  (m_{k},\Sigma_{k}\right )$ as the proposal distribution. In this case, the
acceptance probability of the Metropolis-Hastings algorithm reduces to
$$
\min\left ( \frac{\phi\left
      (g^{-1}(X,\sigma^{2}_{k})(Y-F(X))\right)|g(x^{(l)},\sigma^{2}_{k})| }{
    |g(X,\sigma^{2}_{k})|\phi \left
      (g^{-1}(x^{(l)},\sigma^{2}_{k})(Y-F(x^{(l)}))\right)} ,1\right )
$$
which only depends on the conditional distribution of the observation.

The algorithm to estimate the model parameters can be summarized in the
following scheme:
\begin{enumerate}
\item Start from some initial guess $\Sigma_{0},m_{0},\sigma^{2}_{0}$ and set
  $k=0$ ;
\item Compute $m_{k+1}$ from (\ref{eq:moyenne}) and
  $$
  \sigma^{2}_{k+1} %
  =\frac{1}{N}\sum_{i} \mathbf{E} %
  \left ((Y_{i}-F(X_{i}))'C^{-1}(X_{i})(Y_{i}-F(X_{i})) %
    |Y_{i},m_{k},\Sigma_{k},\sigma^{2}_{k}\right),
  $$
  where $C^{-1}(X_{i})$ is a diagonal matrix whose $j^{th}$ term is
  $(1/(F(X_{i}))_{j})^{2}$.
\item Set $\Sigma_{k}^{(0)}=\Sigma_{k}$ and and set $l=0$
\item for j:=1 to $q=4$\\
  increment $l$\\
  compute $\Sigma_{j,j}^{(l+1)}=S_{opt}+B_{opt}'\left
    (\Sigma_{-j,-j}^{(l)}\right )^{-1}B_{opt}$ and $
  \Sigma_{-j,j}^{(l+1)}=B_{opt}$ where $S_{opt}$ and $B_{opt}$ are
  respectively defined by \eqref{eq:sopt} and \eqref{eq:bopt}~;
\item if $\Sigma_{k}^{(l-4)}$ and $\Sigma_{k}^{(l)}$ are not close enough go
  to step 4. Otherwise set $\Sigma_{k+1}=\Sigma_{k}^{(l)}$ ;
\item Stop if $\left (\Sigma_{k},m_{k},\sigma^{2}_{k}\right )$ and $\left
    (\Sigma_{k+1},m_{k+1},\sigma^{2}_{k+1}\right )$ are close enough.
  Otherwise increment $k$ and go to step 2.
\end{enumerate}
For standard EM algorithms, \emph{i.e.} when no constraint is
imposed on the structure of $\Sigma$, steps 3), 4) and 5) of the
previous algorithm should be replaced by the update of $\Sigma_{k}$
according to equation \eqref{eq:variance}.

It is well known that standard EM algorithms go quickly to a
stationary point of the likelihood during the first iterations and
then take time to converge. Since for stochastic EM algorithms the
criterion being optimized changes randomly at each iteration, it is
somewhat difficult to achieve and check convergence even when the
length $L$ of the simulated Markov Chain is large. Improvements have
been proposed to overcome these problems. In particular Kuhn and
Lavielle \cite{lavielle} suggested updates of the following form~:
$$
\left \{
  \begin{array}{c}
    \Sigma_{new}=(1-\gamma_{k})\Sigma_{k}+\gamma_{k}\Sigma_{k+1} \\
    m_{new}=(1-\gamma_{k})m_{k}+\gamma_{k}m_{k+1} \\
    \sigma^{2}_{new}=(1-\gamma_{k})_{k}\sigma^{2}_{k}+\gamma_{k}\sigma^{2}_{k+1}, \\
  \end{array}
\right.
$$
where $(\gamma_{k})_{k}$ is a decreasing sequence of positive numbers, $\left
  (\Sigma_{k+1},m_{k+1},\sigma^{2}_{k+1}\right )$ is defined as in the
previous algorithm and $\left
(\Sigma_{new},m_{new},\sigma^{2}_{new}\right )$ is the update of
$\left (\Sigma_{k},m_{k},\sigma^{2}_{k}\right )$. Note that this
update scheme forces the algorithm to converge and preserves the PPZ
as well as the positive definiteness of $\Sigma$.

The sequence $(\gamma_{k})_{k}$ should satisfy
$\sum_{k}\gamma_{k}=+\infty$ and $\sum_{k}\gamma_{k}^{2}<+\infty$.
These two conditions are fulfilled when $\gamma_{k}=a/k^{b}$ with
$a>0$ and $b\in(0,1)$. Choosing $\gamma_{k}=a/k$ speeds-up
convergence of the algorithm but the choice of $a$ has to be made
sample by sample. Choosing the same $a$ for all samples can lead to
poor estimations. As practical advice, we suggest choosing
$\gamma_{k}=1/k^{0.8}$. The algorithm takes more time to converge
but a fine tuning of $a$ is unnecessary.

A well known drawback of the EM algorithm is that it does not
produce standard errors as a by-product. We implemented the method
proposed by Jamshidian and Jennrich \cite{jam}. This method relies
on numerical derivation and seems well-suited to the method we
propose. Even if standard errors are helpful for comparing the
results obtained with these two models, the Fisher information
matrix gives pertinent quantitative information only when the sample
size is large enough. However, $N=30$ is probably not a large sample
size. In the next section we use simulations to weight the
performance of the estimation proposed for the cortisol data.

The estimation of the model parameters for the cortisol data
requires some initial estimates to be provided. Thanks to the model
parametrization, we can directly read reasonable values for $m_{0}$
on Figure~\ref{fig:cortisol}. Since the four components of $X$
respectively represent the basal value of cortisol, the maximal
increase, the ``slope" of the sigmoid and the ACTH dose for which
half the maximal increase is obtained we roughly get $m_{0}=\left
  (50,70, 1, 0.1\right )$. We initialize $\Sigma$ with the following diagonal
matrix: $\Sigma_{0}={\rm Diag}\left (0.01 m_{0}^2\right )$. Finally,
for this heteroscedastic model, $\sigma$ can be interpreted as the
coefficient of variation of the cortisol for a given dose. We set it
at $20\%$ that is $\sigma_{0}^{2}=0.2^{2}.$ We estimated $\Sigma$
with EM alone (no constraint was imposed) and with EM+ICF that
preserves the PPZ. In this example, the algorithm seems to converge
in less than 400 iterations. We implemented this algorithm in C++
with a matrix library. Estimates of the parameters obtained with EM
alone were:
$$
\widehat{\Sigma}=\left(
  \begin{array}{cccc}
          21.25 (7.90)&&&\\
          6.28 & 4.25 (1.05)& &\\
          0.13  &0.33&0.047 (0.0071)&\\
          0.015&  0.000867& -0.000267&  0.0000213 (0.000016)\\
  \end{array}
\right),
$$
The figures between brackets are the standard errors for the
variances. We have chosen to give only some standard errors to
lighten the presentation. $\widehat{m}=\left
(50.03,69.81,1.78,0.0845\right )$, $se\left (\widehat{m}\right
)=\left (0.87,1.84,0.085,0.0069\right )$ ,
$\widehat{\sigma^{2}}=0.0145$ and $\log
L(\widehat{\Sigma},\widehat{m},\widehat{\sigma^{2}})=-750.25$.
Estimates of the parameters obtained with the EM+ICF algorithm were:
$$
\widehat{\Sigma}=\left(
  \begin{array}{cccc}
      19.50 (7.85)&&&\\
      -4.66& 2.33(1.12)& &\\
      -0.29  &-0.095&0.058 (0.0063)&\\
      0&  -0.0024& 0&  0.0000144 (0.000012) \\
  \end{array}
\right),
$$
$\quad \widehat{m}=\left (48.84,   71.46, 1.47, 0.0840\right )$,
$se\left (\widehat{m}\right )=\left (0.82,1.81,0.084,0.0071\right )$
and $\widehat{\sigma^{2}}=0.0151$ and $\log
L(\widehat{\Sigma},\widehat{m},\widehat{\sigma^{2}})=-754.23$. We
can see that these likelihoods are about the same and a likelihood
ratio test would not reject the PPZ proposed by the kineticist. The
residual variance estimates are also very close. Surprisingly, there
are quite large differences between the estimates of the third
component of $m$ and the non null components of $\Sigma$.

\begin{rem}[Modelling]
  The general problem of mean and variance modelling for longitudinal data is
  delicate, and several choices are possible, see for instance
  \cite{davidian}, \cite{MR1807760}, \cite{MR1723786,MR1782488},
  \cite{MR1856193}, and \cite{MR2087319}. Our model \eqref{eq:mod2} belongs to
  a standard family of models in PK/PD and was chosen with the kineticist.
  This relatively simple model is heteroscedastic with a constant coefficient of
  variation. An examination of the ``individual'' residuals shows that they
  are centered, which is quite satisfactory.
\end{rem}

\section{Simulations}

The aim of this section is to quantify the potential benefit of
directly estimating a variance matrix with PPZ. We simulated 100
data sets using model \eqref{eq:mod2} with parameters close to the
estimate found in the cortisol data analysis : $N=30$,$m'=\left
(50,70,1.5,0.08\right )$,
$$
\sigma^{2}=0.015\quad\text{and } \Sigma=\left(
  \begin{array}{cccc}
    20 & & &\\
    -4.5 &2.5 &&\\
    -0.3 &-0.1 &0.05 &\\
    0 &-2 \times 10^{-3} &0 &10^{-5}\\
  \end{array}
\right).
$$
Both EM and EM+ICF estimates were calculated. Results are given in Table
\ref{tab:tab1}.

\begin{table}[b]
  \begin{tabular}{lccccccc}
    \hline
    & & & EM& & & EM+ICF& \\
    \hline
    &   True value& Mean&    S.E.&$\sqrt{M.Q.E.}$&   Mean&    S.E.&$\sqrt{M.Q.E.}$\\
    \hline
$\Sigma_{11}$&20 & 17.3 & 10.74 & 11.07 & 18.88 & 9.09 & 9.16\\
$\Sigma_{12}$&-4.5 & -3.64 & 3.16 & 3.28 & -4.1 & 2.39 & 2.42\\
$\Sigma_{22}$&2.5 & 2.27 & 1.48 & 1.5 & 2.74 & 1.25 & 1.28\\
$\Sigma_{13}\times 10^{2}$&-30 & -8.04 & 86.53 & 89.27 & -12.27 & 39.85 & 43.62\\
$\Sigma_{23}\times 10^{2}$&-10 & -7.57 & 19.95 & 20.1 & -10.66 & 14.65 & 14.66\\
$\Sigma_{33}\times 10^{3}$&50 & 74.36 & 95.6 & 98.66 & 73.67 & 65.71 & 69.85\\
$\Sigma_{14}\times 10^{4}$&0 & -7 & 91.62 & 91.89 & 0 & 0 & 0\\
$\Sigma_{24}\times 10^{4}$&-20 & 49.78 & 306.23 & 314.08 & 104.01 & 204.02 & 238.76\\
$\Sigma_{34}\times 10^{5}$&0 & -43.69 & 63.29 & 76.9 & 0 & 0 & 0\\
$\Sigma_{44}\times 10^{6}$&10 & 17.84 & 14.15 & 16.18 & 18.57 & 13.68 & 16.15\\
$m_1$&50 & 51.18 & 1.38 & 1.82 & 48.96 & 1.23 & 1.61\\
$m_2$&70 & 70.73 & 2.60 & 2.70 & 69.52 & 2.44 & 2.48\\
$m_3$&1.5 & 1.54 & 0.22 & 0.22 & 1.49 & 0.21 & 0.22\\
$m_4\times 10^{3}$&80 & 91.44 & 4.08 & 12.15 & 90.06 & 3.87 & 10.78\\
$\sigma^2\times 10^{3}$&15 & 15.91 & 1.77 & 1.99 & 14.07 & 1.57 & 1.82\\
log like.&     & -749.32& 11.39&   & -751.54& 11.68&   \\
    \hline
  \end{tabular}
  \caption{Empirical mean, standard error and square root of mean-quadratic-error of the estimates (M.Q.E.) obtained with EM and EM+ICF.
    The Mean Quadratic-Error is defined as bias$^{2}+$ Variance. }
  \label{tab:tab1}
\end{table}

As expected, the standard errors given in the example are smaller
than those of Table \ref{tab:tab1}. For such sample sizes, which are
often encountered in practice, asymptotic statistics should be
interpreted with care.

The mean parameter $m$ seems to be well estimated. At least on these
simulations, the $\Sigma$ structure influences the estimation of
$m$. However, we notice that the estimates obtained with EM+ICF have
a smaller standard error and mean quadratic error (M.Q.E.) than
those obtained without any constraint. On the whole EM+ICF also
gives estimates with lower bias. This suggests that the mean and
variance estimations are heavily dependent. This sheds light on
approaches, such as the `zero forced'' method, that rely on
estimating the full variance matrix first and modify it by forcing
the PPZ: since all the non zero entries are estimated with the
assumption that the variance matrix does not have prescribed zeros,
they could be poorly estimated. This is consistent with the results
obtained by Ye and Pan \cite{ye} who concluded, in a different
context, that misspecification of the working variance structure
could lead to a large loss in efficiency of the estimators of the
mean parameters.

\begin{figure}[b]
  \begin{center}
    \includegraphics[scale=0.4,angle=-90]{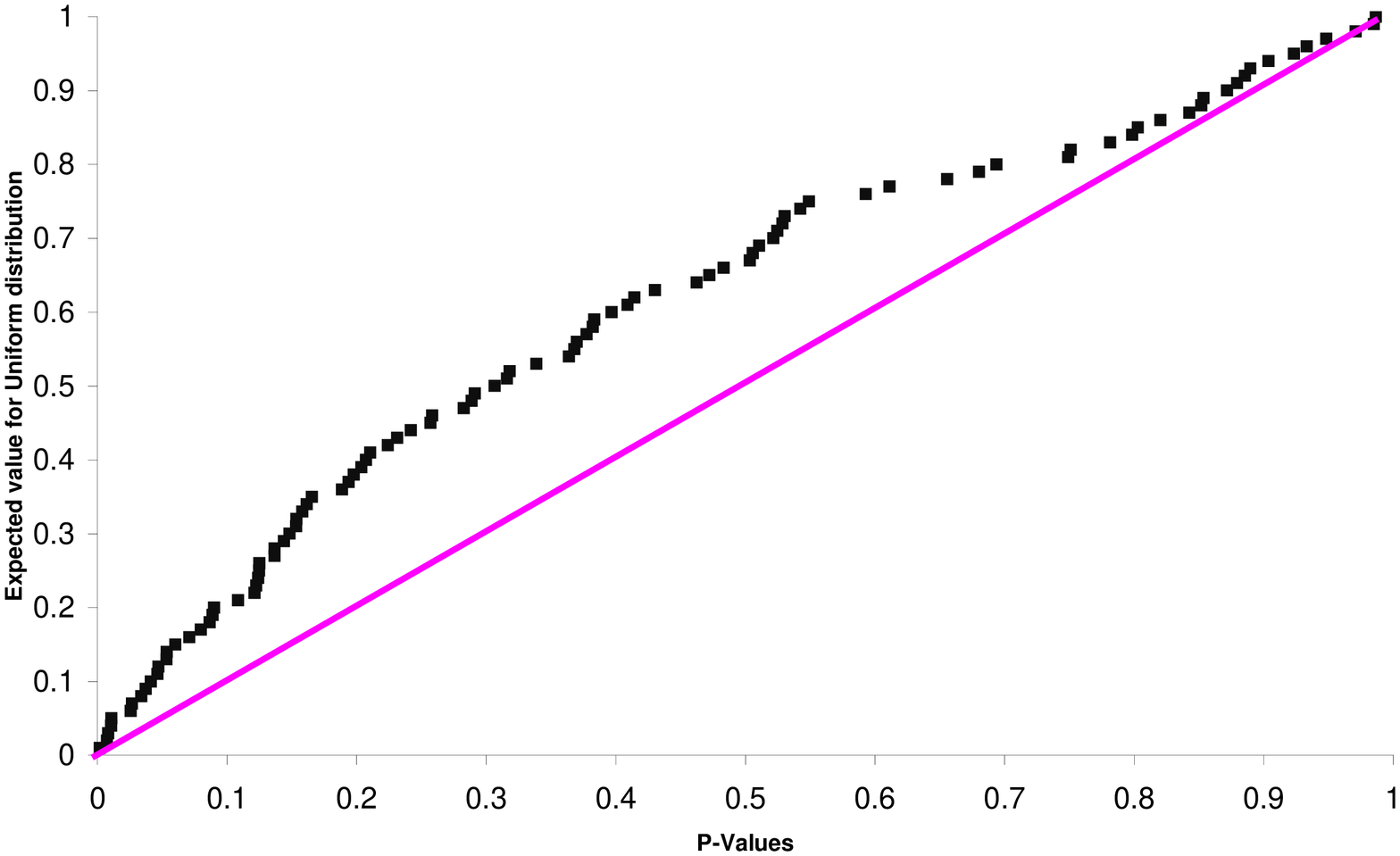}\caption{The Q-Q plot of the
      P-values of the likelihood ratio test versus the uniform distribution on
      $(0,1)$. } \label{fig:QQ}
  \end{center}
\end{figure}

Likelihood ratio tests were performed to test
$$\left \{
 \begin{array}{cl}
   H_{0}:&\Sigma\in S_{4}^{+}(\pi)\\
   H_{1}:&\Sigma\in S_{4}^{+}\\
 \end{array}
\right.
$$
for $\pi=\left \{(1,4); (3,4)\right \}$. Note that whatever the
value of $\pi$, $H_{0}$ is not on the boundary of $S_{4}^{+}$.
Consequently, the likelihood ratio statistics follow asymptotically
a Chi-square distribution under $H_{0}$. Since the data have been
simulated under $H_{0}$, the P-values distribution should be close
to a uniform law on $(0,1)$ at least for large $N$. The Q-Qplot of
the P-Values is represented in Figure \ref{fig:QQ}.
This figure shows that the P-Values are not distributed according to
a uniform distribution and thus the distribution of the likelihood
ratio statistics is not close to a $\chi^{2}$ distribution.
Consequently, $N=30$ is probably not large enough to trust
asymptotic statistics.

\section{Conclusion}
We have proposed a method for the estimation of the variance matrix
with PPZ in nonlinear mixed effects models. This method, which
consists of coupling an ICF like algorithm with an EM like algorithm
gives more efficient estimates than standard EM that ignore the PPZ.
For the sake of simplicity, we have only presented the estimation
algorithm for independent and identically distributed observations.
Extension to different numbers of observations per individual is
straightforward. We also restricted our study to models with
Gaussian $\varepsilon$. More general models in which the
distribution of $\varepsilon$ is not Gaussian and depends on a
parameter $\theta_{2}$ can also be considered. This simply requires
the Metropolis-Hastings chain to be chosen accordingly. We
deliberately chose to show in section 2 a columnwise ICF
implementation that can be extended using theorem 1 to blocks of
$\Sigma$.

Of course, our approach can be adapted without much effort to many
versions of EM and many alternatives to ICF. For pedagogical
reasons, we presented our EM+ICF coupling on a low dimensional
example. The method of course is particularly suited to large
variance matrices with a high percentage of prescribed zero
entries.

\bigskip\textbf{Acknowledgements.} We thank Dr.
Alain~\textsc{Bousquet-M\'elou} who provided the cortisol data set
and for valuable advice regarding the model and the specific
structure of $\Sigma$. We also thank Dr. Mathias~\textsc{Drton} for
interesting discussions on the ICF algorithm and for providing the
last version of \cite{chaudhuri} before publication. The final form
of this article has greatly benefited from the questions and
suggestions of two anonymous reviewers.

\bigskip

\hrule

\bigskip

{\small\noindent
  Didier~\textsc{Concordet}, \url{d.concordet(AT)envt.fr} \\
  Djalil~\textsc{Chafa\"\i}, \url{d.chafai(AT)envt.fr} (corresponding author)
  \medskip\\
  UMR181, INRA, ENVT,\\
  23, Chemin des Capelles, B.P. 87614, F-31076 Toulouse CEDEX 3, France\\
  \textbf{and}\\
  UMR CNRS 5219, Institut de Math\'ematiques de Toulouse, \\
  Universit\'e Paul Sabatier, 118 route de Narbonne, F-31062, Toulouse CEDEX 9, France.}
\newpage

\newpage

\end{document}